\documentclass[twocolumn,aps,showpacs,prl]{revtex4}
\usepackage{graphicx}
\usepackage{dcolumn}
\usepackage{bm}
\def \cuscn{$\kappa$-(ET)$_2$\-Cu\-(SCN)$_2$}

\def \Br{$\kappa$-(ET)$_2$\-Cu\-[N(CN)$_2$]\-Br }

\def \prb{Phys.\ Rev. B}

\begin{document}

\title{The effect of irradiation-induced disorder on the conductivity
  and critical temperature of the organic superconductor $\kappa$-(BEDT-TTF)$_2$\-Cu\-(SCN)$_2$}

\author{James~G.~Analytis,$^1$ Arzhang~Ardavan$^1$,
  Stephen~J.~Blundell$^1$, Robin~L.~Owen$^2$, Elspeth~F.~Garman$^2$,
  Chris Jeynes$^3$ and
Ben~J.~Powell$^4$
}
\affiliation{$^1$University of Oxford, Department of Physics, Clarendon
Laboratory, Parks Road, Oxford OX1 3PU, UK}
\affiliation{$^2$University of Oxford, Department of Biochemistry, South Parks Road, Oxford OX1 3QU, UK}
\affiliation{$^3$University of Surrey Ion Beam Centre,  Guildford GU2 7XH, UK}
\affiliation{$^4$Department of Physics, University of Queensland,
  Brisbane, Queensland 4072, Australia}

\date{\today}

\begin{abstract} 
We have introduced defects into clean samples of the organic
superconductor $\kappa$-(BEDT-TTF)$_2$\-Cu\-(SCN)$_2$ in order to
determine their effect on the temperature dependence of the
conductivity and the critical temperature $T_{\rm
c}$.  We find a violation of Matthiessen's rule that can be
explained by a model of the conductivity
involving a defect-assisted 
interlayer channel which acts in parallel with the band-like 
conductivity.
We observe an
unusual dependence of $T_{\rm c}$
on residual resistivity which is not consistent with the generalised
Abrikosov-Gor'kov theory for an order parameter
with a single component, providing an important constraint
on models of the superconductivity in this material.
\end{abstract}


\pacs{74.70.Kn, 74.62.Dh, 74.25.Fy}

\maketitle

Quasi-two-dimensional organic conductors based on the donor molecule
bisethylenedithiotetrathiafulvalene (BEDT-TTF or ET, see inset to
Fig.~1) have attracted sustained interest \cite{Ishiguro}. These
materials consist of layers of ET separated by inorganic anions and
can be prepared with exceptional purity, so that magnetic
oscillations in the resistivity can be observed at relatively low
fields \cite{goddard}.  Many 
experiments have been conducted to elucidate the symmetry of the
superconducting order parameter 
but despite growing evidence of unconventional superconductivity
\cite{dwave}, the matter has remained unresolved \cite{bjpowell}.

The normal state of these materials is unusual with the
temperature dependence of the electrical transport and the optical
conductivity deviating significantly from what would be expected for
a conventional metal \cite{ross,merino}, showing  a crossover 
from insulating-like conductivity at high temperature to
metallic-like behaviour at lower temperature (see Fig.~1). This
effect has been successfully described within the framework of
dynamical mean-field theory (DMFT) as a crossover from an incoherent
``bad-metal'' state at high temperatures to a coherent Fermi liquid
below $\sim$30~K \cite{ross,merino}.
The usual effect of
disorder is to change only  the temperature independent component of the
resistivity; this is known as Matthiessen's rule. However, several strongly
correlated materials violate Matthiessen's rule including various
cuprate \cite{Cooper} and organic \cite{Lang} superconductors.
Violations of Matthiessen's rule have
previously been taken as evidence for non-Fermi liquid behaviour. In
particular, Strack {\it et al}. \cite{Lang} have argued that the
violations of Matthiessen's rule in the salt \Br indicate that the
description of the conductivity in terms of a crossover from an
incoherent metal to a Fermi liquid is incorrect. 

The behaviour of the superconducting transition temperature $T_{\rm
c}$ as a function of {\it non-magnetic} disorder
\cite{bjpowell,radtketc,andymackenzie} also provides a crucial test
of the symmetry of the order parameter. Anderson's theorem
\cite{anderson} states that for $s$-wave pairing non-magnetic
impurities do not
affect $T_{\rm c}$. For magnetic impurities
$T_{\rm c}$ is strongly reduced for singlet states in a manner
described by the Abrikosov-Gor'kov (AG) formula \cite{abrikosov}.
For non-$s$-wave (including extended-$s$) order parameters, scattering
due to non-magnetic impurities reduces $T_{\rm c}$ in a manner again
described using the AG formula
\cite{bjpowell,radtketc,andymackenzie}.

In order to investigate these effects experimentally for the ET
superconductors, we have selected \cuscn, which has one of the
highest transition temperatures ($T_{\rm c} \approx 10$\,K) in this
family and whose electronic properties have been the subject of
detailed study \cite{goddard}. It is possible to introduce disorder
into these samples by adjusting the cooling rate \cite{stalcup,su},
an effect which is probably due to the freezing of terminal ethylene
group disorder which occurs at around $70$--$80$~K
\cite{tanatar,bjpowell}. However, we find that we have more control
in our experiment by introducing disorder via irradiation by
either x-rays or protons. X-rays and
protons give similar effects, 
but we have been able to perform more detailed experiments
using the former and hence we concentrate mainly on the x-ray damaged
samples
in our discussion. 

In this Letter we report the
violation of Matthiessen's rule in \cuscn. However, we find that a
simple theory which includes the effect of interlayer scattering
from impurities is able to explain this effect. Further we have
confirmed quantitative predictions of this model relating the low
temperature resistivity to the high temperature conductivity. Thus
we find that the violation of Matthiessen's rule is consistent with
the DMFT description of the crossover in
transport behaviour \cite{merino}.  We also observe a disorder-induced
suppression of $T_{\rm c}$ which is consistent with a non-$s$-wave gap.

Our measurements have been performed on single crystals
of \cuscn\ grown electrochemically.
Gold wire of thickness $12.5$\,$\mu$m was attached to the samples
(typical dimensions 0.3$\times$0.3$\times$0.1\,mm$^3$) with graphite
paste in the four-probe configuration. Samples were cooled from
$120$\,K to $10$\,K at a rate of 20\,K/hr, below the slowest rate
used in Ref.~\onlinecite{su}, in order to avoid introducing disorder
from fast-cooling. Defects in our samples were created at room
temperature by filtered Cu $K_\alpha$ radiation ($E=8\,$keV,
$\lambda=1.54$\,\AA) from a Cu x-ray rotating anode (typically
55\,kV, 50\,mA) at a flux of $\approx 2.5\times 10^{8}$ photons s$^{-1}$.
The computer program RADDOSE \cite{raddose} allowed this to
be converted into a dose rate of $\approx 105$\,Gy s$^{-1}$.
Using tabulated
absorption coefficients 
and assuming that the mass absorption coefficient of \cuscn\ can be
calculated as the sum of the mass absorption coefficients of the
constituent atoms \cite{mihaly}, we estimate the x-ray attenuation
length to be $90\ \mu$m, approximately the thickness of the
samples. In order to attain uniform damage we irradiated both sides of
the sample.  X-ray doses up to 630~MGy were used.  Proton
irradiation experiments took place at Surrey Ion Beam Centre. Protons
were accelerated to 4~MeV and could be implanted to a mean depth of
150$\ \mu$m, providing an approximately uniform damage profile for
samples $\leq$100 $\ \mu$m.  For each incremental
radiation dose, we make a
measurement of the transport properties; the contact
configurations stay the same throughout the experiment.  
For each type of irradiation, the
resistivity was found to be reproducible over multiple thermal cycles,
so that we can be confident that the observed changes are due to
irradiation and not to thermal cycling.

It is immediately apparent from Fig.~1 that the effect of increasing
irradiation dose is to {\sl decrease} the resistivity over most of
the temperature range, and in particular to reduce the magnitude of
the broad peak centred around $T_{\rm p}\sim$90~K. Well into the
Fermi-liquid regime however, the behaviour recovers a traditional
metallic character; below $\sim$46~K, the resistivity {\sl
increases} with increasing irradiation dose (see inset to Fig.~1 and
Fig.~2). This violation of Matthiessen's rule is \emph{not}
predicted by DMFT alone for the low defect densities produced in our
experiments and therefore appears at first sight to be at odds with
the DMFT description of electronic transport in layered organic
charge transfer salts.

\begin{figure}
\includegraphics[width=8.2cm]{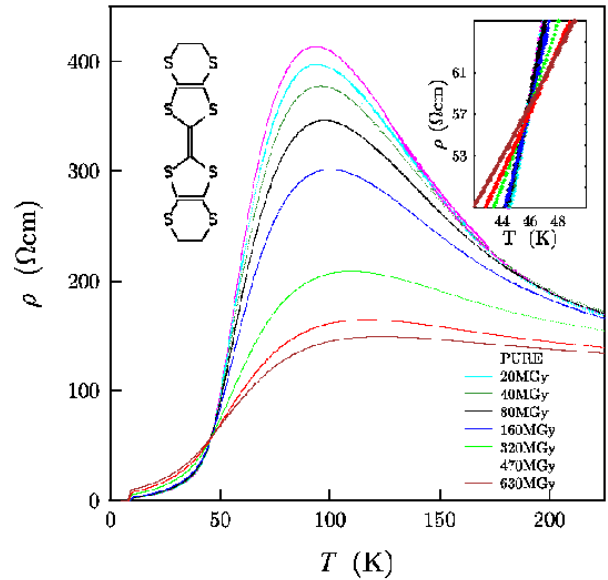}
\caption{Interlayer resistivity of an x-ray
damaged sample as a function of temperature 
for a range of irradiation doses, showing the
violation of Matthiessen's rule. Data from the proton irradiated
samples are qualitatively the same. Insets show the ET molecule and
a magnified view of the region near $T_{\rm cross}$ (defined in the
text).} \label{peakT}
\end{figure}

\begin{figure}
\includegraphics[width=8.2cm]{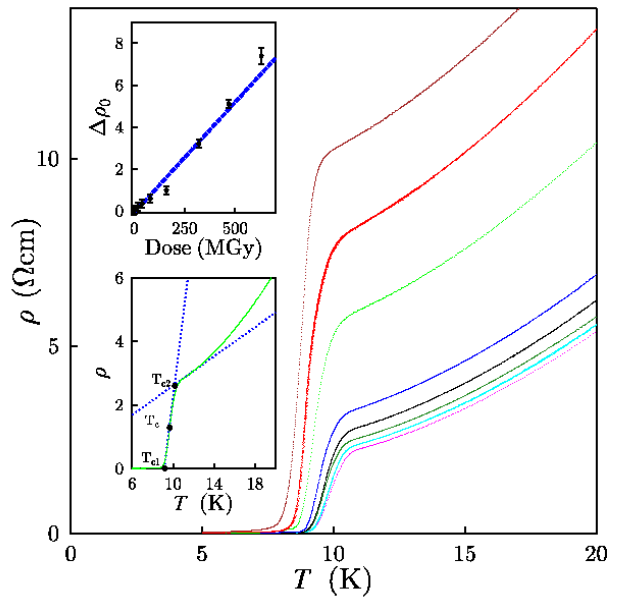}
\caption{The low temperature resistivity of \cuscn\ increases with
x-ray dose in an approximately linear fashion, as shown in the upper
inset. 
Data from
proton-irradiated samples show a similar pattern as a function of
disorder. The lower inset shows three methods of defining the
critical temperature: a lower $T_{c1}$, an upper $T_{c2}$ and
$T_c$, the maximum of the first derivative, which is the
definition used in this paper. 
} \label{lowT}
\end{figure}

The low-temperature region of Fig.~1 is presented in detail in Fig.~2,
more clearly showing the superconducting transition for the same range of x-ray
irradiation doses.  For each trace, the resistivity follows a
Fermi-liquid like $T^2$ dependence at temperatures above the
superconducting transition and a sharp drop at the onset of
superconductivity.  The residual resistivity $\rho_0$ is extracted by
fitting data to $\rho = \rho_0 + AT^2$ in the temperature range from
just above $T_{\rm c}$ to 20\,K.  The upper inset to Fig.~2 presents the
change in residual resistivity $\Delta\rho_0$ with respect to an
undamaged sample as a function of dosage, showing an approximately
linear dependence.

The observation that increasing defect density increases the
interlayer conductivity over a wide temperature range leads us to
hypothesize that defects affect the conductivity in two ways: (i)
the resistivity associated with band-like transport due to the overlap of the
molecular orbitals increases linearly with defect density as
prescribed by Matthiessen's rule; (ii) there is a parallel
defect-assisted interlayer channel, whose conductivity is
proportional to the defect density \cite{interlayer}. This model
suggests that the interlayer conductivity $\sigma(x,T)$ depends on a
dimensionless quantity $x$, which is proportional to the defect
density, and temperature $T$ as
\begin{equation}
\sigma(x,T) = {1 \over \rho_0(0) + x \rho_{\rm imp} + \rho_{\rm intrinsic}(T)}
+ x { \sigma_\perp},
\label{eq:sigma}
\end{equation}
where $\rho_0(0)$ is the residual resistivity of the undamaged
sample, $x \rho_{\rm imp}$ is the contribution to the resistivity
from defect scattering in the transport due to molecular orbital
overlap, $\rho_{\rm intrinsic}(T)$ is the intrinsic
temperature-dependent scattering contribution (due to
electron-electron interactions, phonon scattering, etc.) and $x
\sigma_\perp$ is the defect-assisted interlayer conductivity.

The applicability of this model can be demonstrated by examining low
and high temperature limits. At low temperature $\rho_{\rm
intrinsic}(T)$ is small 
so that
\begin{equation}
\rho(x,T) \approx \rho_0(0) + x \rho_{\rm imp} + \rho_{\rm intrinsic}(T),
\label{lowtres}
\end{equation}
and hence the {\it resistivity} increases linearly with defect
density, $x$. At high temperature $\rho_{\rm intrinsic}(T) \gg
\rho_0(0) + x\rho_{\rm imp} $ and Eqn~(\ref{eq:sigma}) becomes
\begin{equation}
\rho(x,T) \approx \left(x{ \sigma_\perp} +
\rho_{\rm intrinsic}^{-1}(T)
\right)^{-1}
\label{hightres}
\end{equation}
hence the {\it conductivity} $\rho(x,T)^{-1}$ linearly increases
with  $x$, as observed. The inset to Fig.~1 shows that there is a
temperature, $T_{\rm
  cross}$,
at which the resistivity is independent of defect density.  In our
model, this can be found by evaluating ${\rm d}\sigma(x,T)/{\rm
d}x=0$.  We find that a $T$-independent crossing point occurs when 
the conditions (i) $x\rho_{\rm imp} \ll
\rho_0(0)+\rho_{\rm intrinsic}$ and (ii) $\rho(0,T_{\rm cross}) =
(\rho_{\rm imp}/\sigma_\perp)^{1/2}$ are satisfied. For our x-ray
damaged samples, this yields an estimate of $\rho_{\rm
imp}/\sigma_\perp = 3\times 10^3$\,$\Omega^2$cm$^2$. We can obtain
an independent estimate of this parameter by plotting
$\Delta\sigma_{\rm p} = \sigma(x, T_{\rm p}) - \sigma(0, T_{\rm p})$
against $\Delta \rho_0 = \rho_0(x)-\rho_0(0)$, as shown in Fig.~3.
The former quantity can be evaluated in the high temperature limit
(Eqn.~\ref{hightres}) to be $\Delta\sigma_{\rm p} \approx
x\sigma_\perp$, while the latter quantity can be evaluated in the
low-temperature limit to be $\Delta\rho_0 \approx x\rho_{\rm imp}$.
The approximate straight-line dependence (which breaks
down when $\Delta\rho_0$ and $\Delta\sigma_0$ are 
large) observed in Fig.~3 shows
that both quantities are parameterised by $x$ in the same way, and
the gradient of the line yields $\rho_{\rm imp}/\sigma_\perp =
1.5\times 10^3$\,$\Omega^2$cm$^2$, in order of magnitude agreement
with our previous estimate \cite{footnote}. 
We note that for proton irradiation,
both the gradient of the line in Fig.~3 and the temperature $T_{\rm
cross}$ are increased, demonstrating that the nature of the
irradiation affects the ratio $\rho_{\rm imp}/\sigma_\perp$,
suggesting that x-rays and protons produce different types of damage
(protons producing a more effective interplane transport channel
than x-rays).

\begin{figure}
\includegraphics[width=7.6cm]{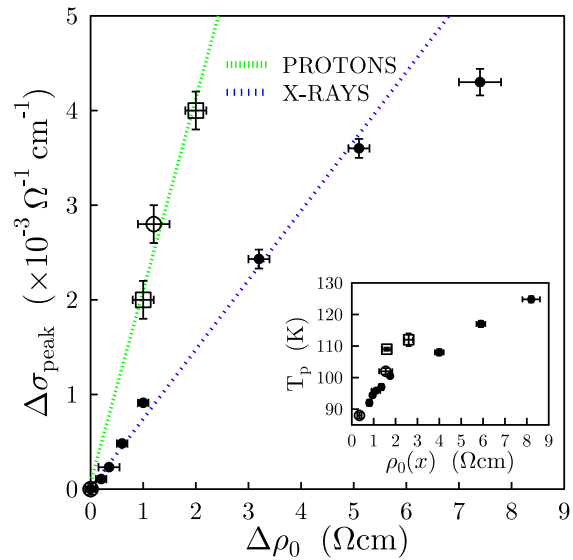}
\caption{The dependence of the change in conductivity at the peak,
$\Delta\sigma$, on the change in residual resistivity,
$\Delta\rho_0$, which is approximately linear, 
except for at large $\Delta\rho_0$.
Due to the different nature of the
defects, the slope of proton (two different samples shown with open
circles and squares) and x-ray (one sample indicated by filled circles)
data differ, though the effect remains qualitatively the same.
Inset: the position $T_{\rm p}$ of the resistivity peak 
shifts slightly to higher temperature with increasing $\rho_0(x)$. }
\label{changeplot}
\end{figure}

\begin{figure}
\includegraphics[width=7.7cm]{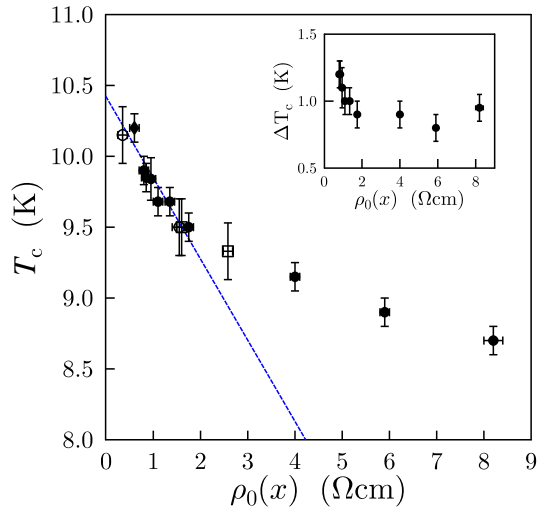}
\caption{The suppression of critical temperature $T_{\rm c}$ compared
to the residual resistivity $\rho_0(x)$.
The suppression is approximately linear for low damage $\rho_0(x)<
\rho_0^{*} = 2\ \Omega$cm, but there is a change in gradient
for $\rho_0>\rho_0^*$. The inset shows the dependence of the
width $\Delta T_{\rm c}$
of the superconducting transition on
$\rho_0(x)$. Symbols correspond to those in Fig.~3 and data for an
additional undamaged sample are shown by a fillled diamond.}
\label{abrikosovres}
\end{figure}

The value of
$T_{\rm c}$, corresponding to
the maximum of ${\rm d}\rho/{\rm d}T$ in Fig.~2, is plotted
as a function of $\rho_0$ in Fig.~4.
We find  that $T_{\rm c}$ falls with defect density
(in agreement with measurements of defects induced by cooling-rate
disorder \cite{stalcup}), but the dependence exhibits a sharp change
in gradient when $\rho_0$ reaches a threshold value
$\rho_0^{*}\approx 2\ \Omega$cm. However, even at the highest defect
densities studied, the samples exhibit a superconducting ground
state. We also find that the {\it width} $\Delta T=T_{{\rm
    c}2}-T_{{\rm c}1}$ (see Fig.\
2) of the superconducting
transition {\it decreases} with increasing defect
density, as shown in the inset to Fig.~4. This is consistent with
$T_{\rm c}$ exhibiting a change of gradient with damage; at high
damage, local variations in damage have a smaller effect on the
broadening of the transition because ${\rm d}T_{\rm c}/{\rm
d}\rho_0$ is lower.

The theory of AG \cite{abrikosov,radtketc} for non-magnetic defects
in a non-$s$-wave superconductor implies that the suppression of
$T_{\rm c}$ follows the AG formula given by
\begin{equation}
{\rm ln}{T_{\rm c}\over T_{{\rm c}0}}=\psi\left({1\over
  2}\right)-\psi\left({1\over 2}+{\hbar\over 
4\pi k_{\rm B}T_{\rm c}\tau}\right)
\label{AG}
\end{equation}
where $\psi$ is a digamma function and $\tau$ is the scattering
time. In the low-defect density limit ($\rho_0<\rho_0^*$), this yields $T_{\rm
c0}-T_{{\rm c}} \simeq \pi\hbar/8k_{\rm B}\tau$. 
This linear sector should have a slope consistent with interlayer
transport theory \cite{bjpowell}, i.e.\
${\rm d}T_{\rm c}/{\rm d}\rho_0=-e^2m^*d_{\perp}t_{\perp}^2 / (4 k_{\rm B}\hbar^3)$,
where $m^*$ is the effective mass, $d_\perp$ is the interlayer
spacing and $t_\perp$ is the interlayer transfer integral. 
Using the value of $m^*$
from transport measurements, 
our
measured slope of the suppression of $T_{\rm c}$ for $\rho_0<\rho_0^*$
yields (the line in Fig.~4), 
a value of
$t_\perp=0.03\pm0.01\ $meV, 
in good agreement with the value
from angle-dependent magnetotransport~\cite{singgod}. Thus the
suppression of $T_{\rm c}$ for $\rho_0<\rho_0^*$
is consistent with a non-$s$-wave gap. 
However, the departure from the AG formula for
$\rho_0 > \rho_0^*$ casts doubt on this interpretation.

We note that
our value of $\rho_0^{*}$
corresponds to
a ratio of the in-plane coherence length to the mean free path
$\xi/l\approx 0.2$.  If all scattering events contributing
to $l$ were pair-breaking, it would be expected that 
superconductivity should be suppressed by this degree of scattering.
Crucially, the observed change of gradient is not expected for a model
involving  an order parameter with a single component
within the 
AG theory.
Even a generalised AG equation describing multicomponent
order parameters \cite{Openov} does
not quantitatively agree with our data, though this approach
assumes
that disorder does not affect the symmetry of the order parameter;
it is probably necessary to examine
specific candidate pairing interactions in detail.
An explanation for this behaviour might involve
a mixed order parameter 
with both $s$-wave and
unconventional components or interband scattering.

In conclusion, we have investigated the effect of radiation-induced
disorder in \cuscn.  We find that a dramatic departure from
Matthiessen's rule can be straightforwardly explained in terms of defect
assisted interlayer tunnelling.  
Although $T_{\rm c}$ initially follows a dependence on 
$\rho_0$ consistent with pair-breaking scattering,
the superconducting state proves to be robust into the dirty limit.
This unusual dependence of $T_{\rm c}$
on $\rho_0$ is not consistent with an order parameter
with a single component, providing an important constraint
on models of the superconductivity in this material.

We acknowledge funding from EPSRC, the Royal Society (A.A.),
BBSCR (R.L.O.) and the ARC (B.J.P.).  We thank
Ross McKenzie, Nigel Hussey and
Paul Goddard for useful discussions and H. Mori for providing
samples. 

\end{document}